\documentclass[aps,twocolumn,superscriptaddress,showpacs,eqsecnum,nofootinbib,pre]{revtex4-1} 
  
\usepackage{scrextend}
\usepackage{amssymb}  
\usepackage{amsmath}   
\usepackage{natbib}
\usepackage{epsfig}   
\usepackage{relsize}
\usepackage{graphicx,subfigure}   
\usepackage{dcolumn}
\usepackage{bm}  
\usepackage[english]{babel}
\usepackage[latin1]{inputenc}
\usepackage{hyperref}
\usepackage{xcolor}
\usepackage{comment}
\usepackage[update]{epstopdf}
\usepackage{appendix}
\usepackage{diagbox}
\usepackage{soul}

\begin{document}
 
\title{Metapopulation dynamics in a complex ecological landscape}

\author{E. H.  Colombo} \email{eduardo.colombo@fis.puc-rio.br}
\affiliation{Departament of Physics, PUC-Rio, Rio de Janeiro, Brazil}
 
\author{C. Anteneodo} \email{celia.fis@puc-rio.br}
\affiliation{Departament of Physics, PUC-Rio, Rio de Janeiro, Brazil}
\affiliation{Institute of Science and Technology for Complex Systems, Rio de Janeiro, Brazil}

\begin{abstract}
We propose a general model to study the interplay between spatial dispersal  
and environment spatiotemporal fluctuations in metapopulation dynamics. 
An ecological landscape of favorable patches is generated like a {\em L\'evy dust}, 
which allows to build a range of patterns, from dispersed to clustered ones. 
Locally, the dynamics is driven by a canonical model for the evolution of the population density, 
consisting  of  a logistic expression plus  multiplicative noises. 
Spatial coupling is introduced by means of two spreading mechanisms:  
diffusive dispersion and selective migration driven by patch suitability. 
We focus on  the long-time  population size  as a function 
 of habitat configurations,   environment fluctuations and coupling schemes.
We obtain the conditions, that the spatial distribution of favorable patches 
and the coupling mechanisms must fulfill, to grant population survival.
The fundamental phenomenon that we observe is the positive feedback between  
environment fluctuations and spatial spread  preventing extinction.
 
\end{abstract}

%% CHECK PACS NUMBERS FOR THIS PAPER !!!!!!!!!!!!!!!!!!!!!!!!!!!!!!!!!!

\maketitle

\section{Introduction}

Habitat fragmentation is commonly observed in nature  associated with heterogeneity in the distribution
 of resources, e.g.,  water, food, shelter sites, physical factors such as light, temperature, moisture,  
and any feature  able to affect the growth rate of the population of a given species~\cite{HanskiBook}. 
A fragmented population made of subpopulations receives in the literature 
the suitable name of \emph{metapopulation}~\cite{HanskiBook,natureHanski1998,levins}. 
These fragments, also known as patches, are not completely isolated as they are coupled due to 
movements of individuals in space. 
For modeling purposes, as a first step one can adopt a  single patch viewpoint, 
taking into account the impact of the surrounding population  in an effective manner~\cite{extinction,migration,otso1,nos2}. 
As a further step  beyond the single patch level, one can resort to a spatially explicit model. 
From this perspective,  deterministic and stochastic  theoretical models have been developed  to obtain the macroscopic behavior 
of the whole population~\cite{hanski,natureHanski1998,natureHanski2000,otso2,otso3,framework}. 
One of the main results is the detection of  critical thresholds that delimit the conditions 
for the sustainability of the population, which  occurs  for a suitable combination of diverse factors, related 
to quality and spatial structure of the habitat, migration strategies and extinction rates.  
Here, we address related fundamental questions in metapopulation theory proposing a model 
that includes a general dispersion process, incorporating  random and selective dispersal strategies. 
Additionally, we investigate the model dynamics on top of  a complex ecological landscape 
whose  spatial structure can be tuned, ranging from spread to aggregated patches.

\begin{figure}[h]
\includegraphics[scale=0.7]{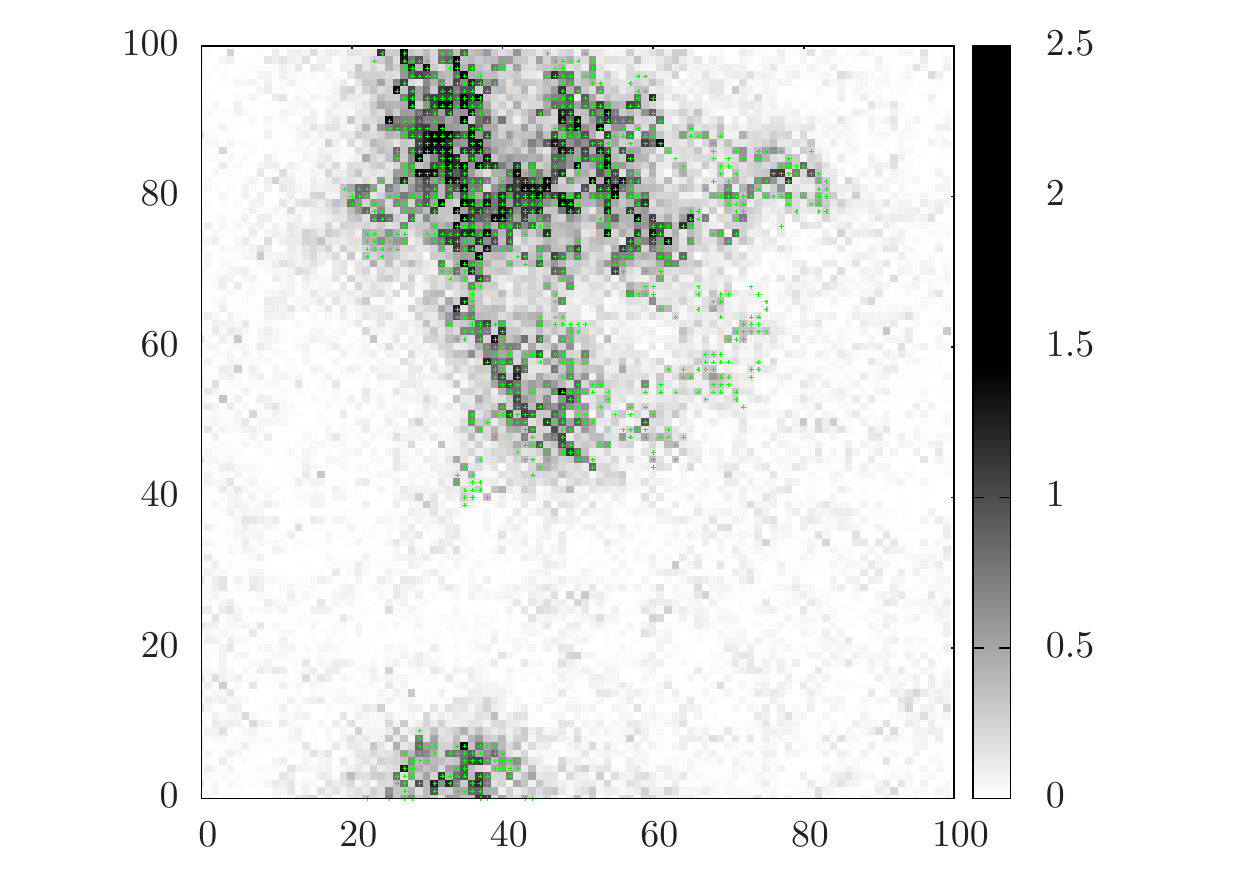}
\caption{Ecological landscape (a green cross  denotes the presence of a favorable patch) and population 
density distribution (in gray scale).}
\label{fig:pattern}
\end{figure}

Let us start from the local dynamics perspective. Locally, each patch 
has its dynamics driven primarily by reproduction and intraspecific competition (carrying capacity). 
Therefore,  we assume that the deterministic factors that rule the evolution of the local population 
can be modeled by the logistic or Verhulst expression~\cite{Verhulst1838}.
Stochasticity is introduced in real systems by the inherent complexity 
of the fluctuating environment (external noise) or by the variations in the birth-death process 
(internal, demographic noise)~\cite{canonical,canonical2,natureHanski1998,nos2}, hence it has to be 
also taken into account. 
These deterministic and stochastic rules, along the lines of the  canonical  
modeling~\cite{extinction,canonical,canonical2}, constitute the local component of our model. 
We assume that this local dynamics takes place on each site of a  lattice, where 
we construct a complex arrangement of favorable and unfavorable patches. 
We define as favorable patches  those sites that induce positive growth at low densities and 
as unfavorable patches  those that are adverse to support life. A typical configuration of the 
model system in a square lattice is illustrated in Fig.~\ref{fig:pattern}.

Spatial coupling is introduced by  migrations from one patch to another. 
First, let us assume that spatial spread is conservative, preserving the number of individuals during travels 
and also that it is nonlocal, in the sense that individuals can travel long distances over the landscape, 
for example like butterflies and birds~\cite{hanski,stanleybirds}. 
We model the populational exchange between patches based on two  behavioral strategies: 
one where the individuals  spread in space diffusively, driven by density differences, 
and another where individuals transit selectively, mainly driven by patch-quality differences.
These strategies can be linked to the amount of spatial information acquired 
by the individuals~\cite{spatialmemory}. 
If they do not have any information about the spatial distribution of favorable patches,  
random movements emerge. In fact, this has been the focus of works on  animal foraging, 
where optimal efficiency in  resource  search occurs without previous knowledge of food distribution~\cite{fractal}. 
This type of behavior has isotropy as a main trait, indicating directional indifference. 
Instead, if individuals have information on the distribution of favorable patches, 
some directions will be preferred.

In Sec.~\ref{sec:model}, we will describe in detail each part of the model.
The spreading process and spatial configuration of the ecological landscape are described in 
Secs.~\ref{sec:spread} and~\ref{sec:landscape}, respectively. 
The  results,  reported in Secs.~\ref{sec:deterministic} and~\ref{sec:stochastic},  
focus on the impact of the spatial arrangement of the habitat  
on its  overall viability, that is, on the long-time behavior of the population size. 
Mainly numerically, and with the aid of analytical considerations, we investigate the impact of  habitat topology, 
spread range and stochasticity  in the long time behavior of the population size, 
 compared to the corresponding uncoupled metapopulation.

\section{Model}
\label{sec:model}

In mathematical terms, we assume that the evolution of the population density (number of individuals per unit area),  
$u_i$, in each patch $i$ is described by
\begin{equation}
\dot{u}_i = a_iu_i -bu_i^2 + D\, \Gamma_i[u] + \sigma_\eta u_i \bullet \eta_i(t) + \sigma_\xi \sqrt{u_i}\circ \xi_i(t) \, ,
\label{maineq}
\end{equation}
where the constants $a_i$, $b$, $D$ are the  growth rate, 
  the intraspecific competition coefficient and the spatial coupling coefficient. 
The noises $\eta_i$ and $\xi_i$ introduce  environment   and demographic stochasticities, respectively. 
They are assumed to be mutually independent zero mean and unit variance Gaussian white noises.
The intensities of these multiplicative noises are controlled by coefficients $\sigma_\eta$ and $\sigma_\xi$. 
The environmental noise term  is expected to have external origins, then, its correlation even if small is non-null, 
justifying the use of Stratonovich calculus ($\bullet$); 
at the same time, the demographic noise represents fluctuations in the reproduction process of each independent individual, 
then It\^o calculus is more suitable ($\circ$)~\cite{canonical,extinction,vankampen}. 
These contributions define a local dynamics, at each site, ruled by Eq.~(\ref{maineq}) with $D=0$, 
which is known as canonical model~\cite{extinction}.
We incorporate the additional term $D\Gamma_i[u]$ into Eq.~(\ref{maineq}) to account 
for the nonlocal contribution arising from fluxes between  patches, as it will be explained below. 
This is the term that couples  the set of stochastic differential equations (\ref{maineq}).

The intrinsic growth rate of each patch $i$ is quantified by  the growth rate  $a_i$, 
that can take positive or negative values.
Patches can be favorable (or not), promoting growth (or decrease) of the local population, 
with $a_i = A_i^+>0$ (or $a_i = A_i^-<0$). 
For the sake of simplicity, we consider a binary landscape, where
sites can be in any of  two states,   $A_i^+=-A_i^-=A>0$,   
as assumed in previous studies~\cite{kraenkel,kraenkel2}.

\subsection{Nonlocal  coupling}   
\label{sec:spread}

 In order to define the coupling scheme  let us state some considerations. 
First, note that it is reasonable to assume that active individuals like butterflies, 
birds, terrestrial animals use their perception and memory to increase the efficiency in the search for viable habitats. 
The  spatial information stored by the individuals can yield  optimized routes between favorable regions. 
In fact, there is a relation between spatial memory and migration strategy~\cite{spatialmemory}. 
We introduce this trait by allowing individuals to have access to information about the spatial distribution of patch quality. 
Spatial knowledge can be acquired, for instance,  by a direct verification in a previous visit 
or by the perception of the collective dynamics.  
Otherwise, if individuals do not have  any information about the ecological landscape, or if they do not have memory, 
 uncorrelated trajectories can emerge.
 
We contemplate both scenarios by modeling spread through a diffusive component together 
with a contribution of direct routes connecting favorable patches, governed by  quality  differences.
The relative contribution of both mechanisms is regulated by parameter $\delta$, with $0\le \delta \le 1$ 
tuning from the ecologically driven ($\delta=0$) to the purely diffusive ($\delta=1$) cases.   
Moreover, we assume that coupling is weighted by a factor $\gamma(d_{ij})$ that decays with 
the  distance $d_{ij}$ between patches $i$ and $j$, as will be defined below. 
Then,  the flux $J_{ij}$ from patch $i$ to $j$ is given by 
\begin{equation}
J_{ij} = \left[\delta + (1-\delta)\alpha_{ij} \right]\gamma(d_{ij})u_i \geq 0\, ,
\label{flow}
\end{equation}
where $\alpha_{ij} \equiv (a_j-a_i)/(4A) + 1/2$. Hence, the total flux   is
\begin{align}
\Gamma_i[u] &  = \mathlarger\sum_{j\neq i} (J_{ji} - J_{ij}) \notag \\
&= \mathlarger{\sum}_{j} \gamma(d_{ij})\left[\delta(u_j-u_i) + (1-\delta)(\alpha_{ji}u_j-\alpha_{ij}u_i)\right]  \, .
\label{gamma}
\end{align}
The total density  is conserved by the exchanges described by Eq.~(\ref{gamma}), as can be seen by summing over $i$. 
It indicates that individuals  tend to move towards patches with fewer individuals and  better quality. 
For $\delta=1$, Eq.~(\ref{gamma}) represents a generalization of the Fick's law for  nonlocal dispersal driven by 
density gradients. 
For $\delta=0$, with our definition of $\alpha_{ij}$, and binary patch growth rate, 
the possible values of $ \alpha_{ji}u_j-\alpha_{ij}u_i $ are
 \begin{center}
\begin{tabular}{|c|c|c|}
\hline 
 \backslashbox{$j$}{$i$} & $A$ & $-A$   \\
\hline
$A$ & $(u_j-u_i)/2$  &   $u_j$    \\
\hline
$-A$ &  $-u_i$ & $(u_j-u_i)/2$     \\
\hline
\end{tabular}
 \end{center}
This means that, when the quality of two patches is different, the flux occurs in the direction of the higher quality, 
weighted by the out-flowing population density (lowest quality patch). 
Only when the quality is the same, diffusive exchange can occur, to allow a network of favorable patches. 
 
Concerning the factor that takes into account the distance between patches, 
there is empirical evidence~\cite{hanski,dispersal} that the frequency of occurrence of flights between 
patches decays with the distance, which is reasonable due to the increase of energetic cost. 
Although diverse decay laws are possible, 
we will assume exponential decay  of the weight $\gamma$ with the traveled distance $\ell$, 
as observed for some kinds of butterflies~\cite{hanski,dispersal,dispersal2}, that is
\begin{equation}
\gamma(\ell) = \mathcal{N}^{-1} \exp(-\ell/\ell_c)\, ,
\end{equation}
where $\ell_c$ is a characteristic length (the average traveled distance) and the normalization 
constant $\mathcal{N}$ is such that the sum of the contributions  of all patches  equals  one.   
Operationally, we will truncate the exponential at $\ell \simeq 8\ell_c << L$, where $L$ is the linear characteristic 
size of the landscape.

\subsection{Ecological landscape}
\label{sec:landscape}

In nature, the arrangement of the ecological landscape is built by many distinct processes, occurring in many time scales, 
 creating  complex spatiotemporal structures. 
Then, beyond the inclusion of the environmental noise $\eta$, it is also important 
to take into account the spatial organization of patches~\cite{natureHanski2000,hanski,spatialMeta,spatialMeta2}. 
  
\begin{figure}[h!]
\includegraphics[width=\linewidth]{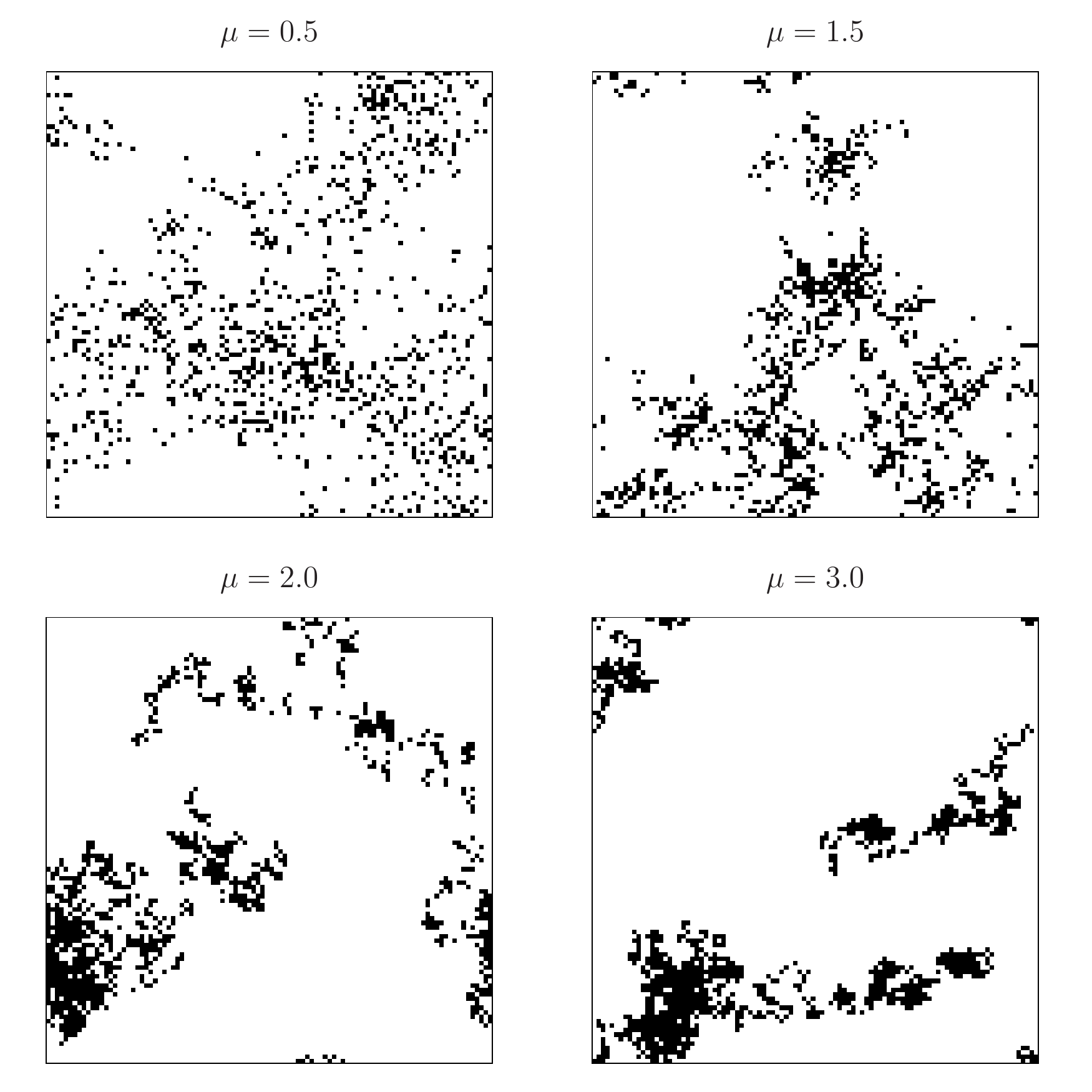}
\caption{Habitat topology for different values of the exponent $\mu$. 
Black cells indicate positive growth rate $A$ and white cells  negative growth $-A$,  
in a square domain of linear size $L=100$. 
The density of favorable patches is $\rho = 0.1$.
}
\label{fig:habitats}
\end{figure}

Heterogeneity and patchiness are adequate to capture the complexity of diverse ecological 
systems~\cite{landscape,levyland1,levyland2,ecofractal1,ecofractal2}.
Here we propose to use as complex ecological landscape a {\em L\'evy dust}~\cite{fractal} 
distribution of favorable patches on a square domain of size $L\times L$ patches, with periodic boundary conditions. 
Over a background of adverse patches ($a_i=-A$), we construct a L\'evy dust of favorable patches 
($a_i=A$) given by the sites visited by a L\'evy random walk  with step lengths $\ell$ drawn from the probability density function 
$p(\ell) \propto  1/\ell^\mu$, with $1 \le \ell \le L$. 
This protocol has been used in the  study of 
different problems~\cite{fractal,levyland1,levyland2}, but we apply it 
here in the study of metapopulation dynamics. 
It allows to mimic a general class of realistic conditions~\cite{landscape,levyland2,ecofractal1,ecofractal2} 
and   to tune  different habitat landscapes through parameter $\mu$, from widely spread (for $\mu=0$) to 
compactly aggregated in a few clusters separated by large empty spaces (for $\mu=3$), 
as illustrated in Fig.~\ref{fig:habitats}.

\begin{figure}[b!]
\includegraphics[width=\linewidth]{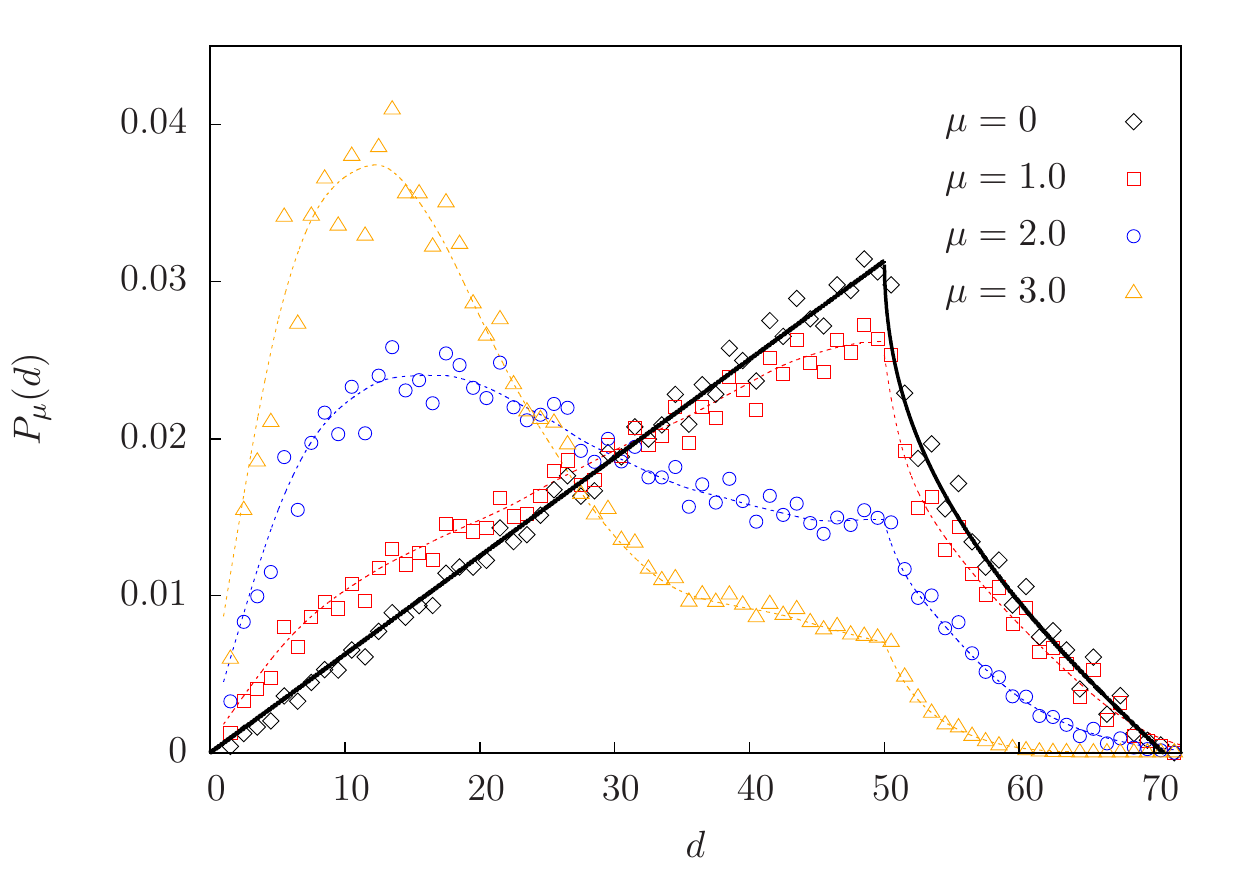}
\caption{Probability distribution of the {\em minimal} distance between favorable patches 
for different values of $\mu$,  for $\rho=0.1$ and $L=100$ (100 configuration were used). 
Fluctuations are due to the discrete nature of the possible distances in the lattice. 
The dotted lines are a guide to the eye.  
The solid line represents the  probability distribution for the distance between 
uniformly distributed random points in continuous space, drawn for comparison.}
\label{fig:distance}
\end{figure}

We quantify the change in the spatial structure by computing  
the  probability distribution of the {\em minimal} distance $d$ between favorable patches $P_\mu(d)$ 
(see Fig.~\ref{fig:distance}). 
For the density $\rho=0.1$ used in the figure, 
when  $\mu\lesssim  1$,  patches are typically  far from each other. 
For high values ($\mu\gtrsim 3$), the generating walk  approaches the standard random walk, 
creating a much more clustered structure, evidenced by the peak at short distances. 
However the shape of  $P_\mu(d)$ changes with $\rho$. 
When the patch density $\rho$ is high,  the shape of $P_\mu(d)$ resembles that of the uniform arrangement even for large $\mu$,  
while  at low densities   $P_\mu(d)$ presents a peak at small $d$ since the resulting configuration of patches 
is very localized even for small $\mu$, as will be discussed in Sec.~\ref{sec:density}. 
Furthermore, $P_\mu(d)$ is also sensitive to $L$,    
but we kept $L$ fixed ($L=100$), even if some properties may have not attained the 
large size limit, as far as $\mu$ and $\rho$ allow 
to scan many qualitatively different possibilities of landscape structure.

Concerning the factor $\gamma_\mu$ that reflects the topology, as defined in Eq.~(\ref{gammamu}), 
 it is affected by $\rho$ more through  the amount of favorable patches $n_v$ than by its indirect consequences 
on the spatial distribution $P_\mu$.

\subsection{General considerations about the model}
\label{sec:general}

The set of parameters $\{D, \delta, \ell_c   \}$ regulate the nonlocal dynamics. 
While $D$ is the strength of the nonlocal coupling,   $\delta$ controls the balance between 
diffusion and directed migration, and  $\ell_c$ defines  the coupling range. 
The ecological landscape is characterized by $\mu$ and $\rho$.
 
In the results  presented in the following sections, we will restrict  the analysis 
to a region of parameter space relevant  to discuss the main phenomenology of the model.
Thus, we will set  $A=b=1$ in all cases. 
We will also consider $L=100$ and typically $\rho=0.1$. 
Concerning the noise parameters,  we set  $\sigma_\eta=\sigma_\xi=0$ to analyze the 
deterministic case in Sec.~\ref{sec:deterministic} and turn noise on   by setting 
$\sigma_\eta=\sigma_\xi=1$ in Sec.~\ref{sec:stochastic}.  This choice is based on  
previous works~\cite{canonical,extinction}. Indeed, population size 
can be subject to large fluctuations as demonstrated by experimental data~\cite{stanleybirds}.  
 
We performed numerical simulations of  Eq.~(\ref{maineq}) on top of different 
landscapes, by preparing  the system in 
the stationary state of the deterministic and uncoupled case, 
 i.e., $u_i(0) = \max\{a_i/b,0\}$ for all $i$, plus a small noise. 
Integration of Eq.~(\ref{maineq}) was carried out with Euler-Maruyama scheme with a time step $\Delta t=10^{-3}$.

\section{Deterministic case ($\sigma_\eta=\sigma_\xi=0$)}
\label{sec:deterministic}

Before proceeding to study the full model, we consider the deterministic case. 
Locally, when stochastic contributions are neglected, 
the asymptotic value of the population size for each patch is  $u_i =a_i/b$. 
Introducing  nonlocal effects, the population size might change. 
If population exchanges between patches are guided solely by their quality   ($\delta=0$), 
then, the favorable-patch network will conserve the initial population size, 
so no interesting phenomena occur from the viewpoint of extinction. 
However, when $\delta>0$, the diffusive behavior induces exploration of the neighborhood 
independently of  habitat quality, which leads to the occupation of  unfavorable regions 
making likely the death of individuals.

\begin{figure}[h!]
\includegraphics[width=1\linewidth]{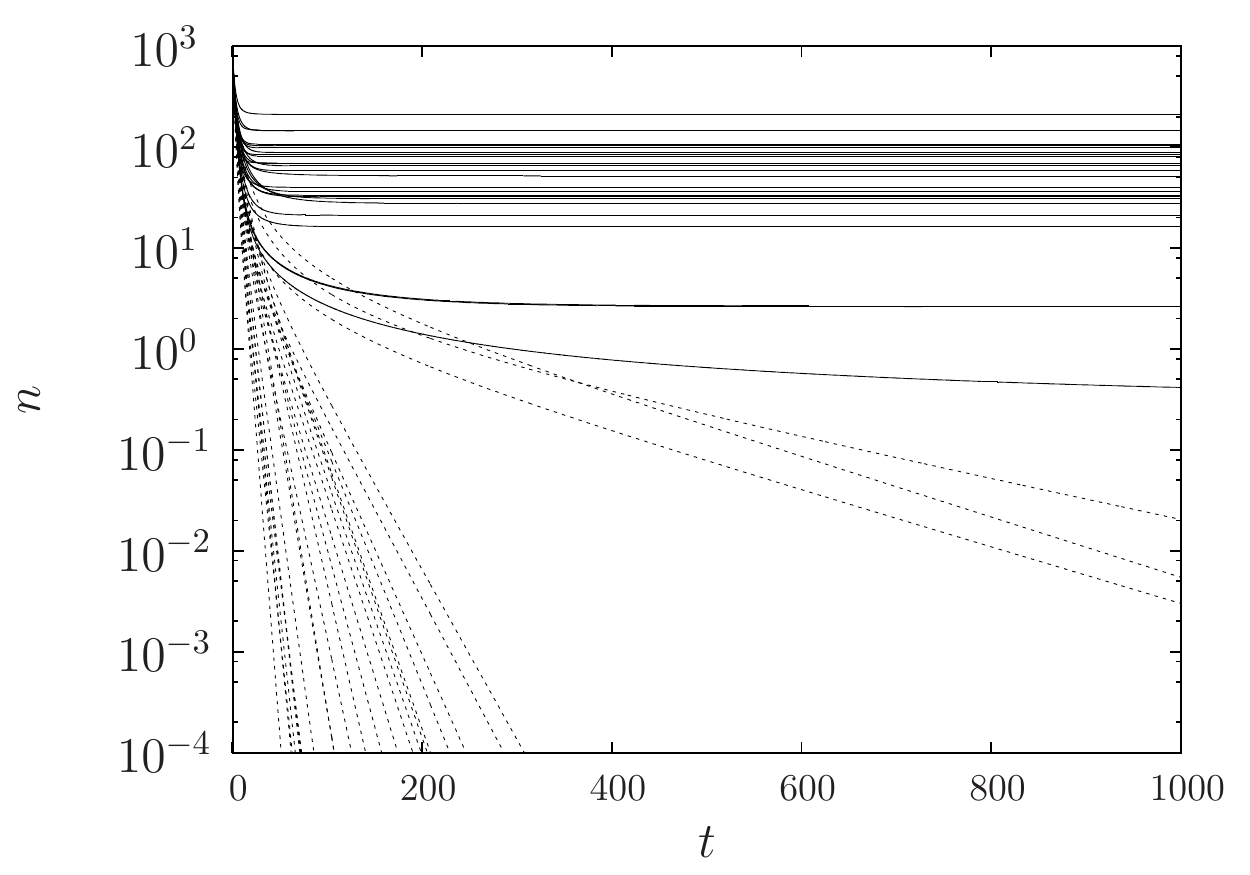}
\caption{Deterministic ($\sigma_\eta=\sigma_\xi=0$)  time evolution 
 of the total population density $n$,  
for $ \delta=1$, $\rho=0.1$,   $D=10$, $\ell_c=0.5$, $\mu=1.7$, and different initial landscapes.  
This set of values results in about half of 50 realizations leading to extinction. 
We use a dotted line to flag the ones that tend to extinction exponentially fast 
and a solid line for those that lead to population survival.}
\label{fig:single}
\end{figure}

By numerical integration of Eq.~(\ref{maineq}) we obtain 
the time evolution of the total population density $n(t)=\sum_{i=1}^{L^2} u_i(t)$. 
In Fig. \ref{fig:single} we show  the outcomes   
for fixed values of the model parameters and different initial conditions (different landscapes). 
While some of the realizations lead to exponential  decay of the population  
other ones attain finite values at long times. 
Several different non null steady states can be attained.
Notice however, that  the steady values of different realizations are all 
below that of the uncoupled case, $\rho L^2 A/b =1000$ for the parameters of the figure.
Hence, diffusion favors the decrease of the total population density and the occurrence of extinctions, 
as expected.

\begin{figure}[b!]
\includegraphics[width=\linewidth]{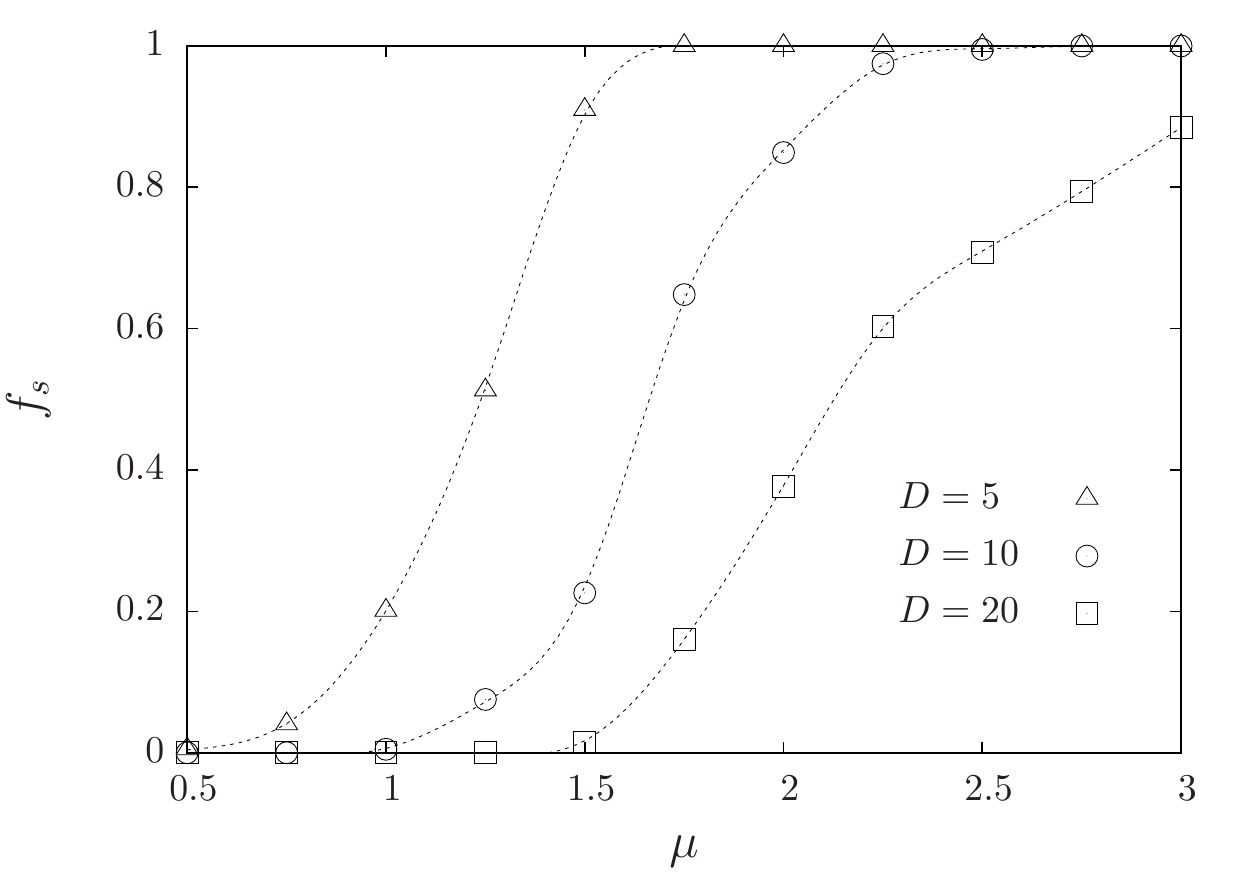}
\caption{Fraction of surviving metapopulations  $f_s$ (over 100 realizations) in the deterministic case 
($\sigma_\eta=\sigma_\xi=0$)  as a function of $\mu$, 
setting $ \delta=1$, $\rho=0.1$, 
$\ell_c=0.5$, for the values of $D$ indicated on the figure. 
In this and following figures, dotted lines are a guide to the eye. }
\label{fig:det}
\end{figure}

In order to investigate how the fraction of survivals changes with the topology, we plot 
in Fig.~\ref{fig:det}   the  number of survivals per realization, $f_s$, as function of $\mu$, 
for several values of $D$. 
Besides the initial condition used throughout this paper (see Sec.~\ref{sec:general}), 
we observed that a perturbation of the null state also leads to the same results of Fig.~\ref{fig:det}. 
For given $\mu$,  increasing $D$ favors the occurrence of extinctions as already commented above.
For given $D$, below a threshold value of $\mu$ the population gets  extincted in all the realizations, 
while above a second threshold it always survives (for the finite number of realizations done), 
between thresholds  both states, the null and non null ones, are accessible. 
The number of   non null stable states   increases with $\mu$.

Summing over all $i$  the deterministic form of Eq.~(\ref{maineq}), 
one finds that the steady solution  $\dot{n}=0$ must satisfy $\sum_i A_i u_i=b\sum u_i^2$, which has 
infinite solutions between the fundamental null state and the uncoupled case solution (the only stable one for $D=0$). 
The condition for stationarity of the total density depends only on the local parameters, 
since fluxes are only internal, 
however,  the coupling and landscape can stabilize configurations other than the trivial ones.  
Furthermore, in the Appendix, we performed an approximate calculation  to show that, for small $D$, 
the null state   is stable if 
  \begin{equation}
  A\,-\,D(1  - \gamma_\mu) >0\, ,
	\label{thrshld1}
\end{equation}
where $0 \le \gamma_\mu\le 1$ is a factor that mirrors the topology, 
varying from $\gamma_\mu=\rho$ for the uniform case $\mu=0$ to $\gamma_\mu=1$ in the limits of large $\mu$ or large $\rho$.
% (see Fig.~\ref{fig:rhos}).   
 Despite this approximate expression fails  in providing  accurate threshold values, it predicts that survival 
is facilitated by larger  $A$ and spoiled by increasing $D$.
It also qualitatively predicts the impact of the topology, as far as it indicates 
that the destructive role of diffusion can be compensated by a large enough  degree of 
clusterization of the resources given by large $\gamma_\mu$.

\section{Stochastic case}
\label{sec:stochastic}

\begin{figure}[h!]
\includegraphics[width=\linewidth]{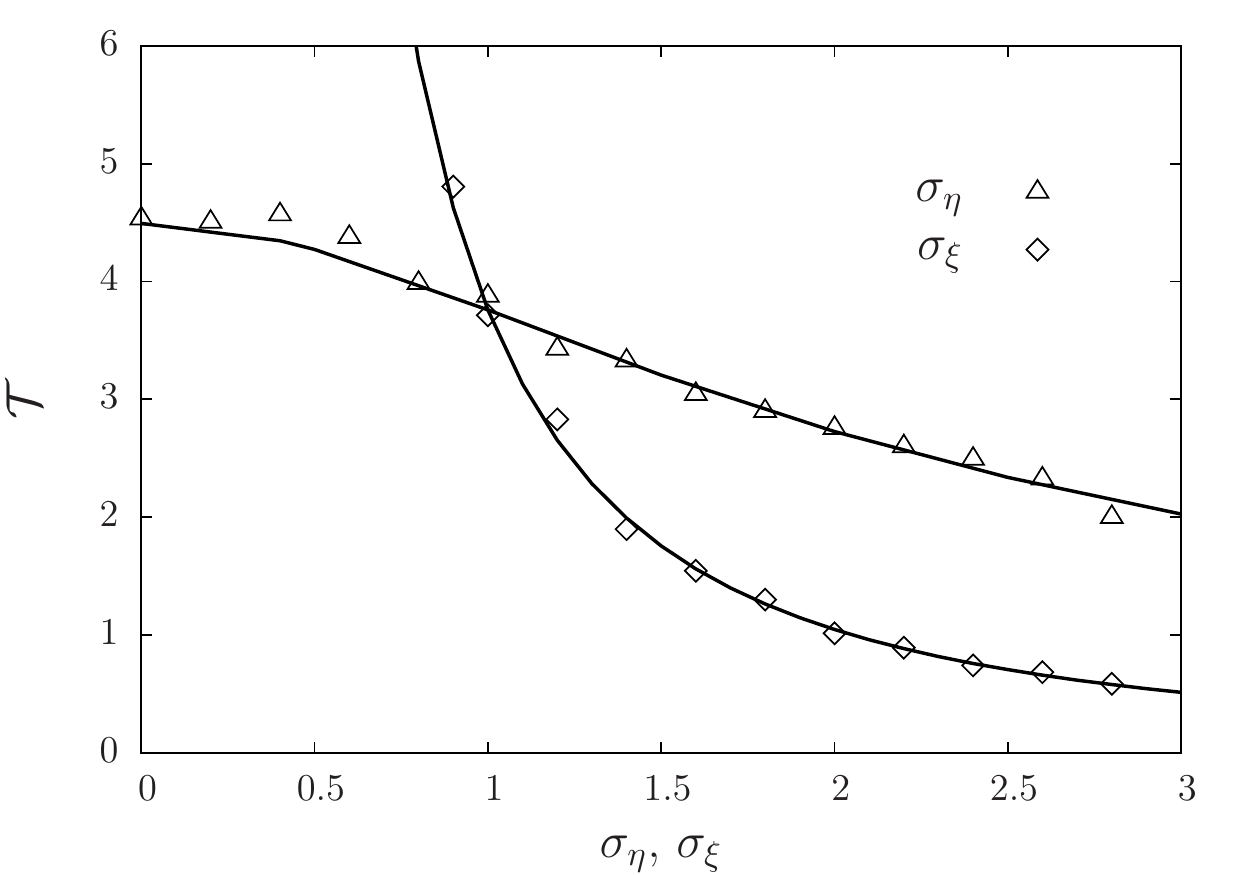}
\caption{Single patch dynamics. Mean extinction {\cal T} time vs $\sigma_\xi$ (for $\sigma_\eta=1$) and 
vs $\sigma_\eta$ (for $\sigma_\xi=1$).  
Symbols correspond to numerical simulations averaged over 500 samples 
and the full lines    to the theoretical prediction given by Eq.~(\ref{mte}). 
The curve for variable $\sigma_\xi$ diverges  in the limit  $\sigma_\xi\to 0$. 
}
\label{fig:mte_sx1}
\end{figure}

First let us review some known results about the local (one site) dynamics, 
which is obtained in the limit $D\to 0$ of  Eq.~\ref{maineq} (canonical model). 
In the deterministic case, the two-state  habitat~\cite{kraenkel,kraenkel2} 
leads to local extinction (if $a_i=-A$) or   finite population (if $a_i=+A$). 
The presence of stochastic contributions changes  the stability of the patches.  
When $a_i=-A<0$,  the local extinction event predicted deterministically is reinforced by noise.  
For $a_i=+A>0$, the demographic noise $\xi$,     %in the presence of noise $\eta$,  
leads to extinction in a finite time  that diverges as $\sigma_\xi \to 0$\cite{otso1,extinction}.  
The external noise $\eta$ reduces the most probable value of the population size,   
that becomes very close to zero when  $\sigma_\eta > \sqrt{2A/b}$~\cite{horsthemke2006noise}.

The population stability can be quantified by the mean time to extinction $\mathcal{T}$  
averaged over realizations starting at $u(0)$. 
For Eq.~(\ref{maineq}) with $D=0$, $\mathcal{T}$ is  given by~\cite{canonical},

\begin{equation}
\mathcal{T} = \int_0^{u_0} \int_z^\infty \frac{\exp\left(\int_z^v \Psi(u) du\right) }{V(v)}dvdz \, ,
\label{mte}
\end{equation}
where $\Psi(u) = 2M(u)/V(u)$, with $M(u) = au-bu^2 + \sigma_\eta^2u/2$ and $V(u) = \sigma_\eta^2 u^2 + \sigma_\xi^2 u$. 
The results of Eq.~(\ref{mte}) are in good accord with those from numerical simulations, as illustrated in Fig.~\ref{fig:mte_sx1}. 
When the noise intensity decreases, the time to extinction always increases, being divergent  in the limit $\sigma_\xi\to 0$.

\subsection{From local to global behavior}

\begin{figure}[b!]
\includegraphics[width=\linewidth]{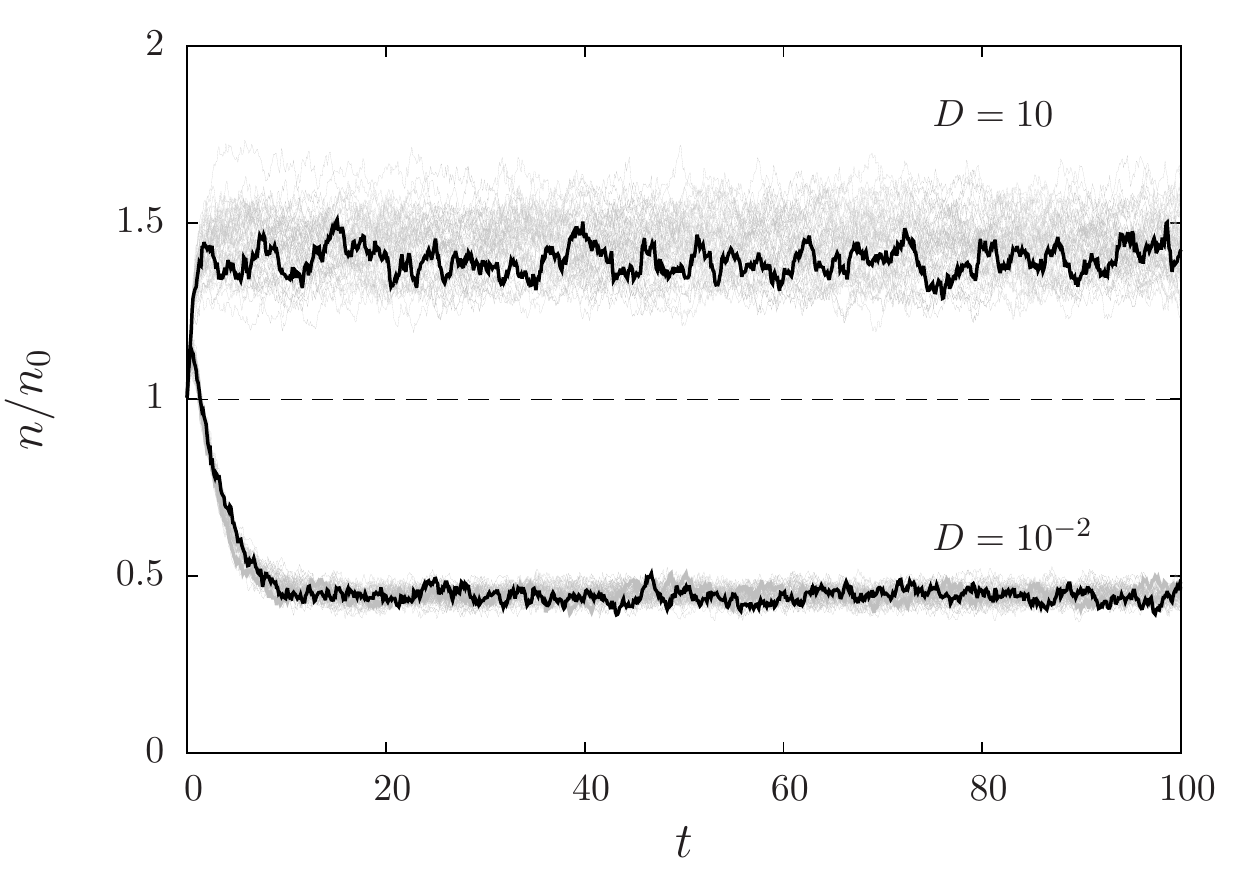}
\caption{Temporal evolution of $n/n_0$  for 
  $\delta=0.5$, $\rho=0.1$,       $\ell_c=0.5$, $\mu=2.0$, $\sigma_\eta=\sigma_\xi=1$ 
and values of $D$ indicated on the figure.  
We highlight a single realization (black full line) for each set of 50 realizations (gray lines). 
The dashed line at $n=n_0$ is plotted for comparison. 
}
\label{fig:evolution}
\end{figure}

In this section we investigate the effects introduced by patch coupling, i.e., when $D\neq 0$. 
Nonlocal contributions redistribute the individuals  in space, 
driven by density and quality gradients. 
In Fig. \ref{fig:evolution} we show  that  $D\neq 0$ prevents the extinction events that occur when
  $D=0$ (see Fig.~\ref{fig:mte_sx1}). Therefore, in contrast to the deterministic case, now spatial coupling 
	is constructive.
On the other hand, noise has also a constructive role when $D\neq 0$, differently to the   uncoupled case,   
not only preventing 
extinction  but also contributing to the increase of the population (as in the case $D=10$). 
In a previous work~\cite{nos2}, we already observed the constructive role  in population growth 
of linearly multiplicative Stratonovich noise in contrast with the destructive behavior of its  It\^o version.  
 Therefore, environmental noise and coupling have a positive feedback effect on  population growth, 
as shown in Fig.~\ref{fig:evolution}. 

We will compute the long-time  total population density $n_\infty \equiv \lim_{t\to \infty} n(t)$, which is 
useful to be compared with the initial value  $n_0 \equiv n(0)= \rho L^2 u_0 = \rho L^2 (A/b)$, 
that  represents the asymptotic total density in the deterministic uncoupled case. 
Then we will measure the ratio $E \equiv \langle n_\infty \rangle /n_0$, that represents a kind of efficiency, where 
the brackets indicate average over landscapes and noise realizations.

\begin{figure}[b!]
\includegraphics[width=\linewidth]{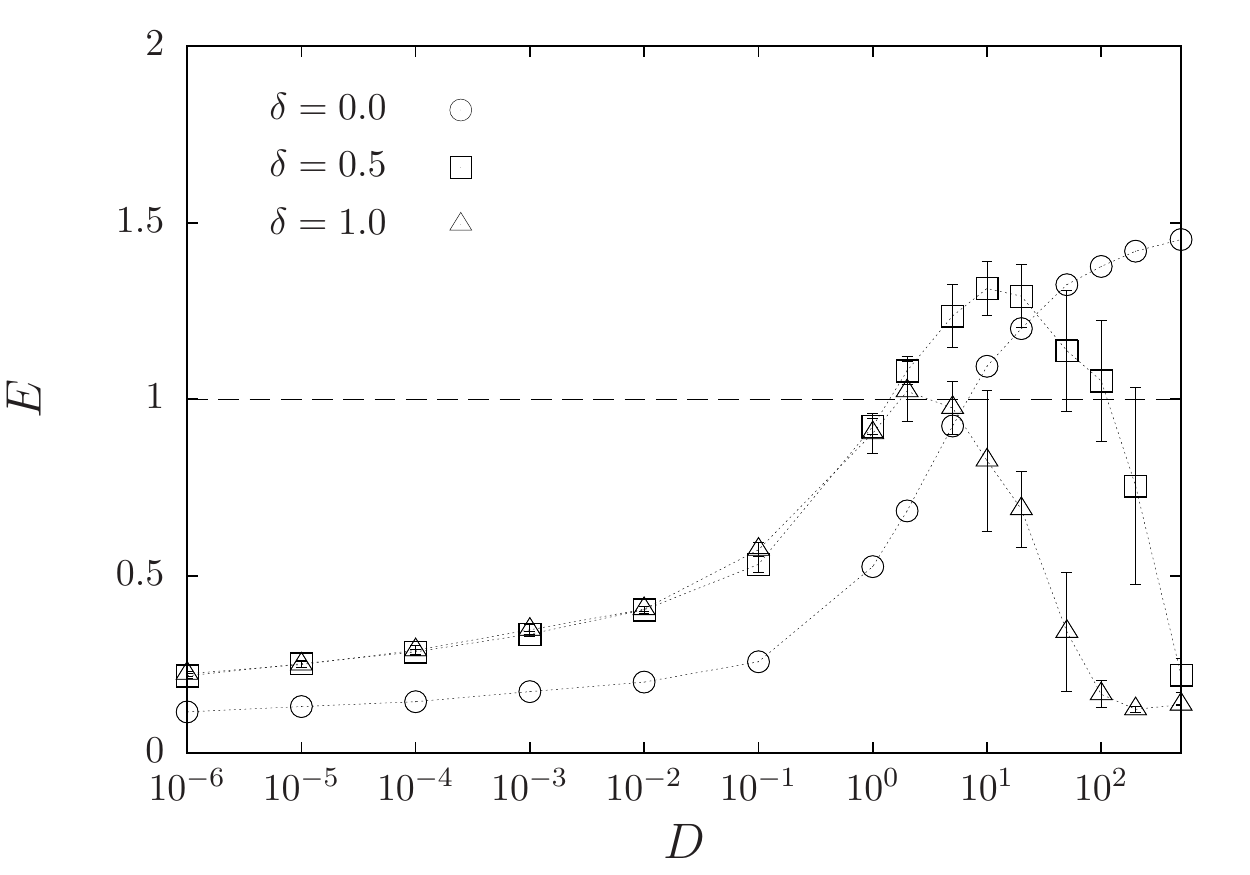}
\includegraphics[width=\linewidth]{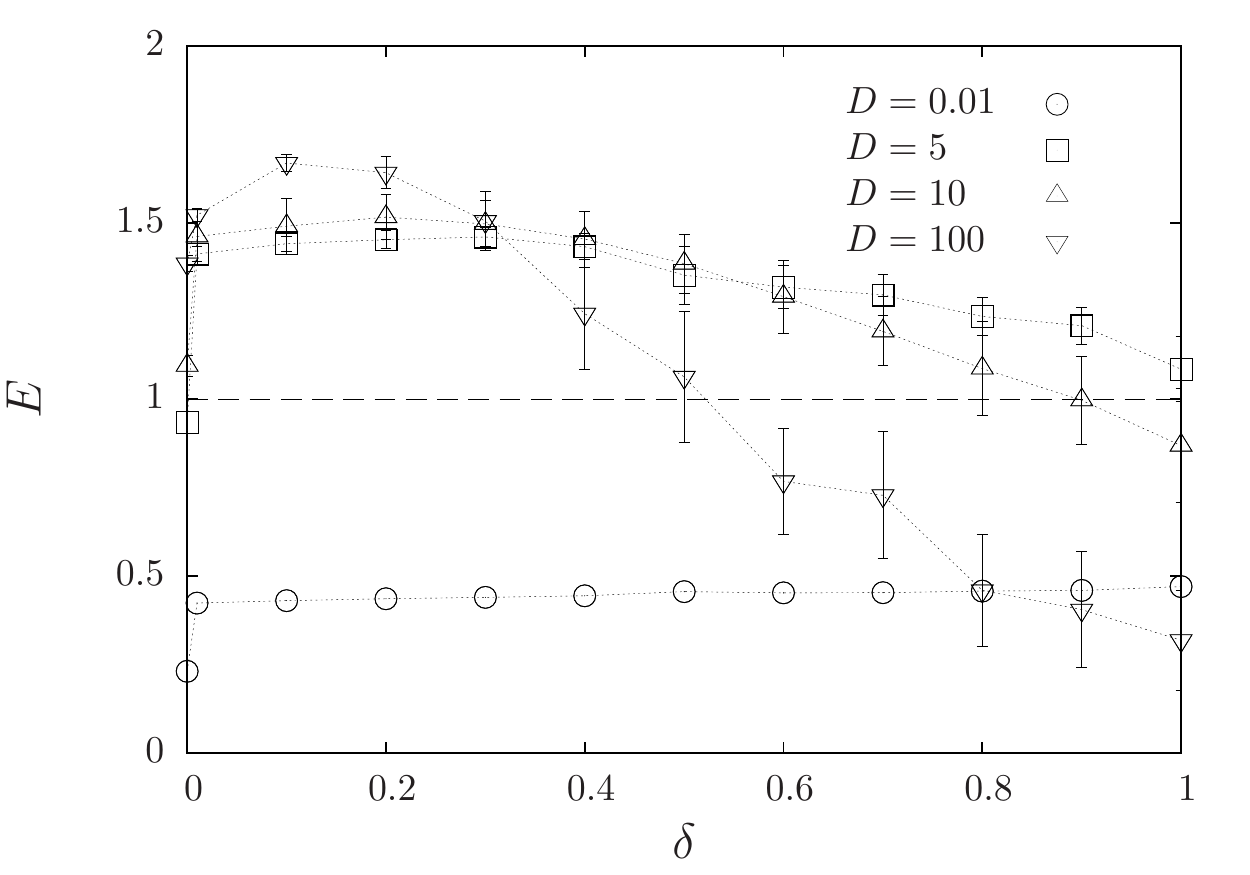}
\caption{Ratio $E\equiv \langle n_\infty \rangle /n_0$ as a function of $D$ (upper panel) 
for different values of $\delta$, and $E$ as a function of $\delta$, for different values of $D$ (lower panel), 
and    $\rho=0.1$,     $\ell_c=0.5$, $\mu=2.0$ and $\sigma_\eta=\sigma_\xi=1$. 
The symbols represent the average over 20 samples and the vertical bars the standard error.
The dashed line at $E=1$ is plotted for comparison.
}
\label{fig:D_mte}
\end{figure}

In the upper panel of Fig.~\ref{fig:D_mte} we plot   the ratio $E$ as a function of $D$.
We see that for very small values of $D$, the population is 
non null, although the ratio $E$ is smaller than one. Moreover, for given $D$, 
the ratio $E$ is smaller when the diffusive component is absent ($\delta=0$).
In all cases the ratio first increases with $D$ and even exceeds the value $E=1$, 
indicating again that not only the noise has a constructive role in preventing extinction but also in promoting 
the increase of the initial total population.  
When diffusion is present ($\delta>0$), the increase of $E$ occurs up to an optimal value of the coupling $D$ 
(with $E>1$), above which the ratio decays. 
Hence, there is a nonlinear effect that does not reflect the linear combination in 
Eq.~(\ref{flow}), as shown in the lower panel of Fig.~\ref{fig:D_mte}. 
The diffusive component, despite being much less efficient, like in  placing individuals in unfavorable regions, 
 acts with greater connectivity. 
Then, for small $D$, the nonlocal contribution of the diffusive coupling is much higher than in the $\delta=0$ case, 
leading to a higher population size. 
In fact, the abrupt transition in the connectivity of the spatial coupling is mirrored in  the abrupt change 
suffered by the ratio $E$   as $\delta$ becomes non null.
Contrarily, for high values of $D$, $\delta=0$ is more efficient due to high damage caused by an intense 
dispersal towards unfavorable regions, which in the case of Fig. \ref{fig:D_mte} are the majority of the sites.
All these observations  highlight  the importance of the diffusive strategy, that can become more efficient than the 
ecological pressure driven by the quality gradient.

\subsection{Habitat topology and coupling range}

\begin{figure}[h]
\includegraphics[width=\linewidth]{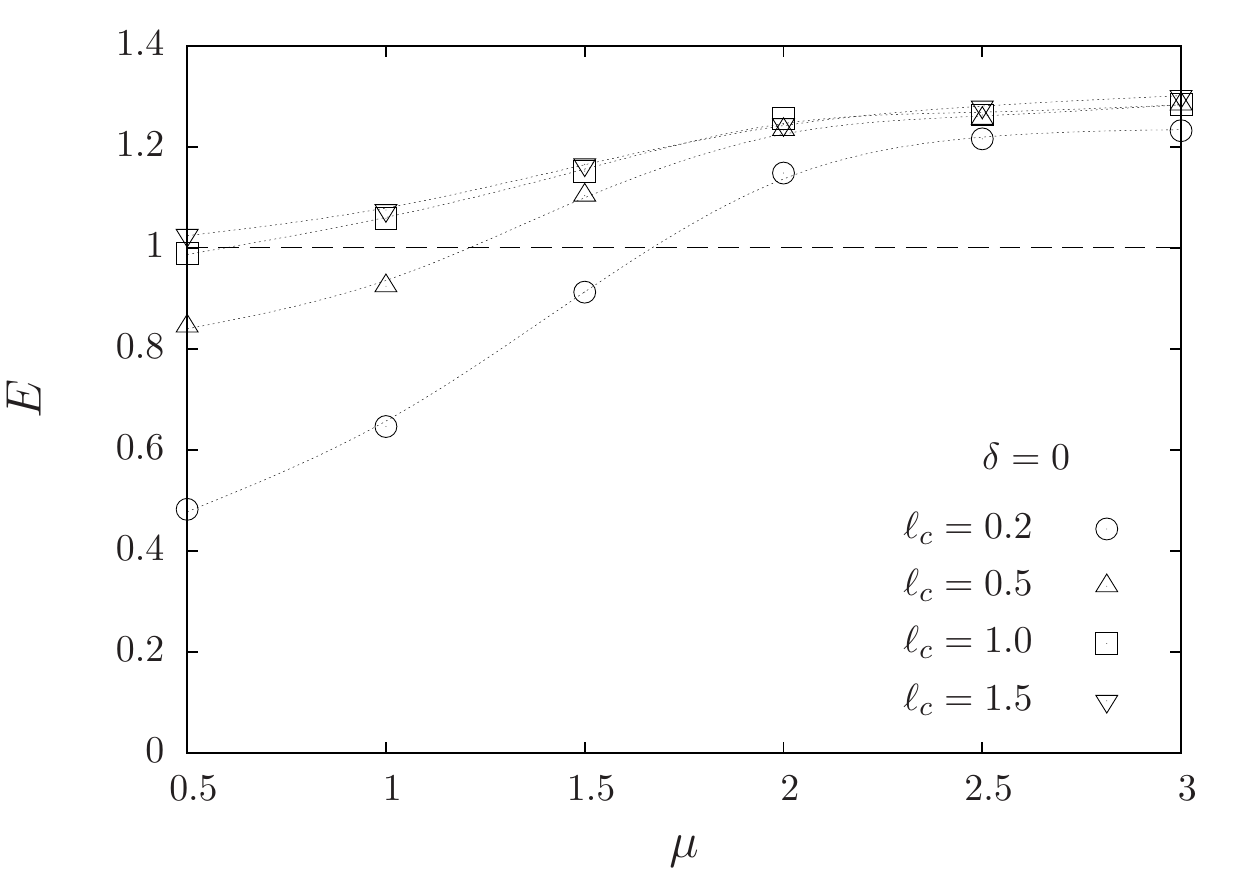}
\includegraphics[width=\linewidth]{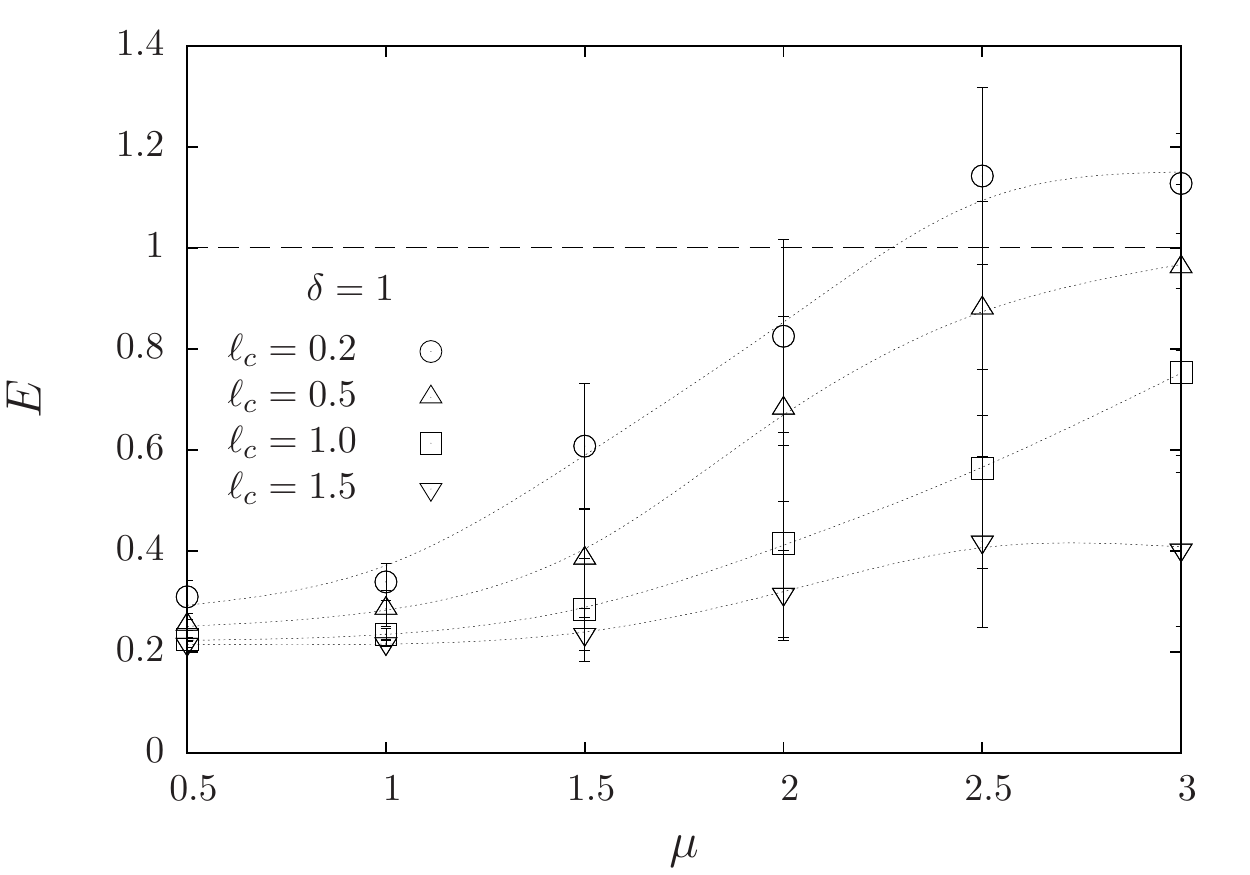}
\caption{ Ratio $E\equiv \langle n_\infty \rangle/n_0$, as a function of $\mu$ for different values of $\ell_c$, 
when $\delta=0$ (upper panel) and $\delta=1$ (lower panel), with $\rho=0.1$, $D=20$ and $\sigma_\eta=\sigma_\xi=1$. 
The symbols represent the average over 20 samples and the vertical bars the standard error.
The dashed line at $E=1$ is plotted for comparison. 
}
\label{fig:phase}
\end{figure}

The nonlocal contribution results from the combination of the spread strategies,  
interaction range  and topology, characterized by $\delta$, $\ell_c$ and $\mu$, 
respectively. Fig.~\ref{fig:phase} shows the ratio $E$ as function of $\mu$ with different 
values of $\ell_c$ for $\delta=1$ and $\delta=0$.

$E>1$ means that the combination of  habitat topology and spatial coupling range 
leads the population to profit from the environment fluctuations, increasing its size. 
The  region $E>1$ is bigger when individuals are  selective with respect to their destinations 
($\delta=0$) and increases with $\ell_c$.
For the diffusive strategy ($\delta=1$), $E>1$ is attained only in a clustered habitat (large $\mu$) 
together with  short-range dispersal (small $\ell_c$). We have already seen that in a sparse habitat, 
diffusion  represents a waste, specially if the dispersal is  long-range.
Instead, when $\delta=0$, the habitat does not need to be so clustered or the 
range so short for population growth. 
In this instance, the optimal combination occurs in a clustered habitat but with long-range coupling. 
Finally note that, as $\ell_c$ increases, $E$ becomes independent of the topology.

\subsection{Density of favorable patches}
\label{sec:density}
 
\begin{figure}[h]
\includegraphics[scale=0.63]{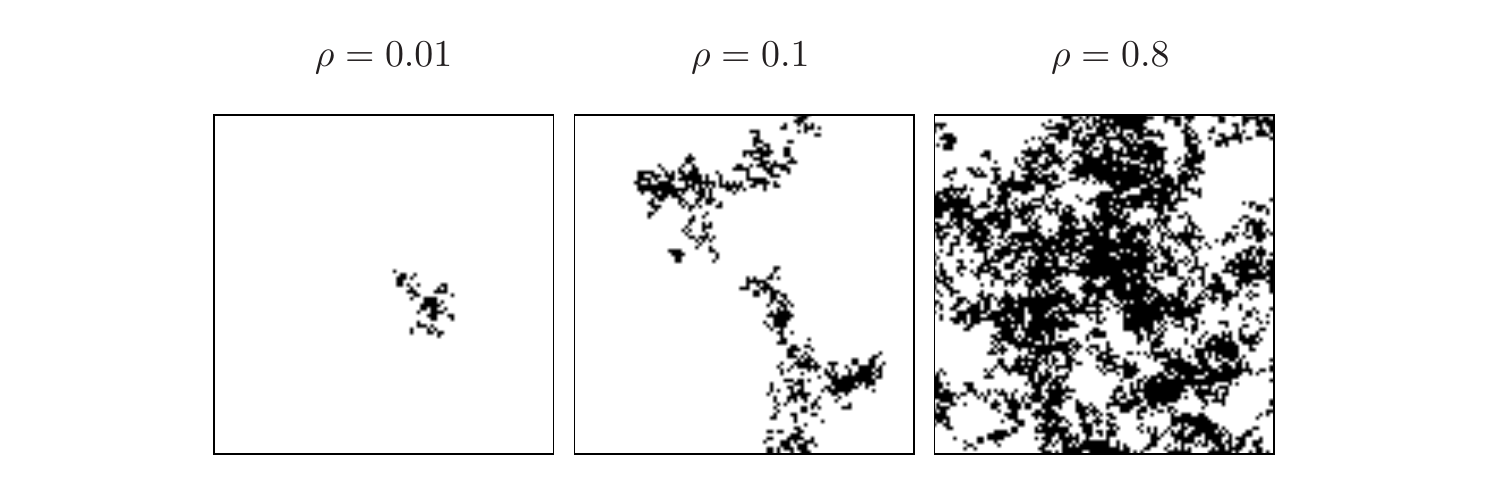}
\includegraphics[width=\linewidth]{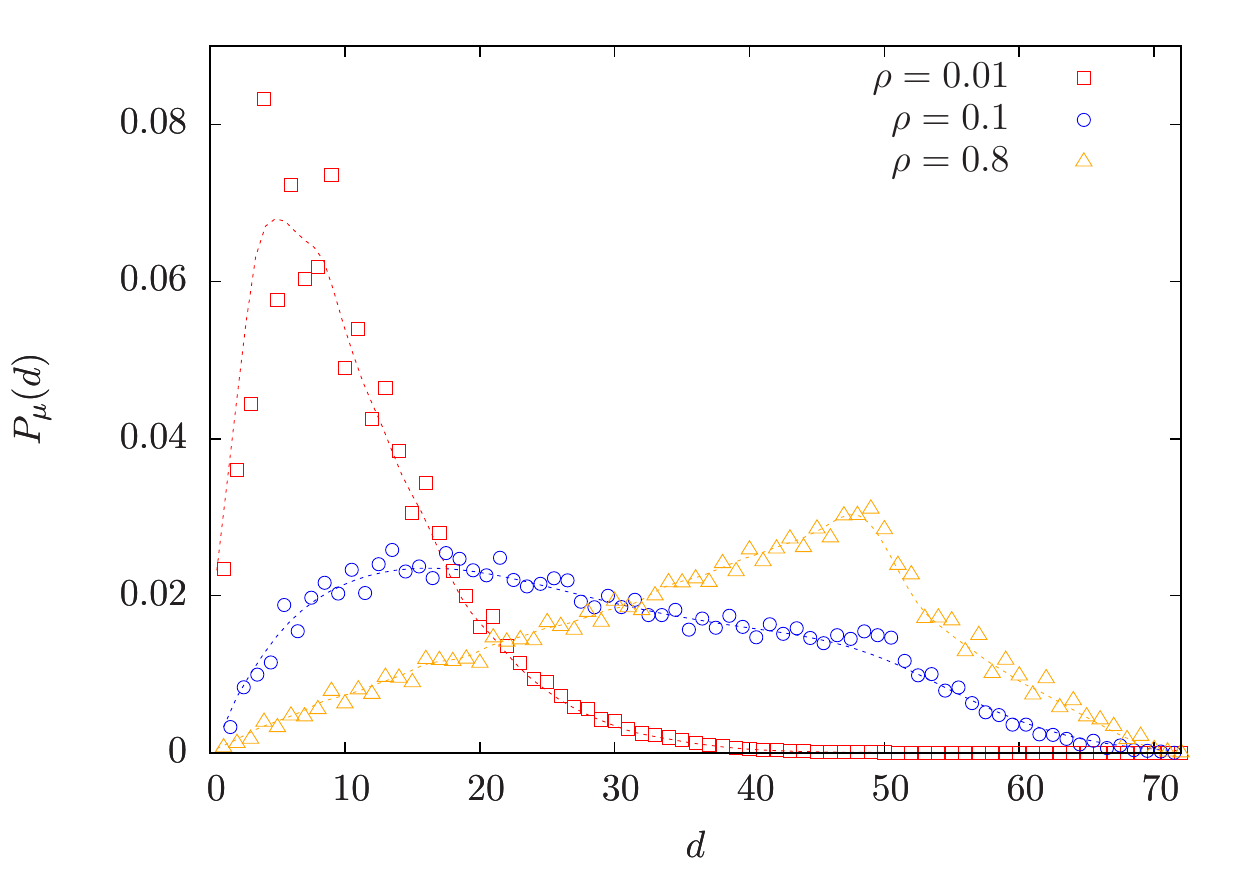}
\caption{Spatial structure and probability distribution of the minimal distance between favorable patches 
(averaged over 100 landscapes), for  $\mu=2$ and three different values of $\rho$ indicated on the figure.}
\label{fig:rhos}
\end{figure}

Another  important issue is the influence of the density $\rho$ of favorable patches in the dynamics. 
Until now, we have kept it constant to highlight the effects of the heterogeneity of the habitat and 
of the coupling schemes in the longtime behavior of the total population size. 
In terms of the protocol used to generate the ecological landscape,  
 $\rho$ not only changes  the proportion of favorable patches but also reshapes the distribution of distances  
between favorable patches. 
 In Fig.~\ref{fig:rhos} we show three different outcomes of the spatial structure and the corresponding 
distance distribution for a fixed value of $\mu=2$. 
For low $\rho$, patches organize in a kind of archipelago structure, that is much smaller than the system size, and 
the distance resembles that obtained for large $\mu$ when $\rho=0.1$.
For high $\rho$, many points of the domain are visited creating a distance distribution that approaches the homogeneous form. 
For  $\mu$ higher than the value of the figure,  profiles very similar to those shown in Fig.~\ref{fig:rhos} are obtained. 
Meanwhile,  for small values of $\mu$,  the distribution 
is almost invariant with $\rho$, being very close to that of the uniform case. 
This is due to frequent flights with lengths of the order of system size.
 Concerning the factor $\gamma_\mu$ that reflects the topology, as defined in Eq.~(\ref{gammamu}), 
 it can be affected by $\rho$ more through  the amount of favorable patches $n_v$ than by its indirect consequences 
on the spatial distribution $P_\mu$.

\begin{figure}[h]
\includegraphics[width=\linewidth]{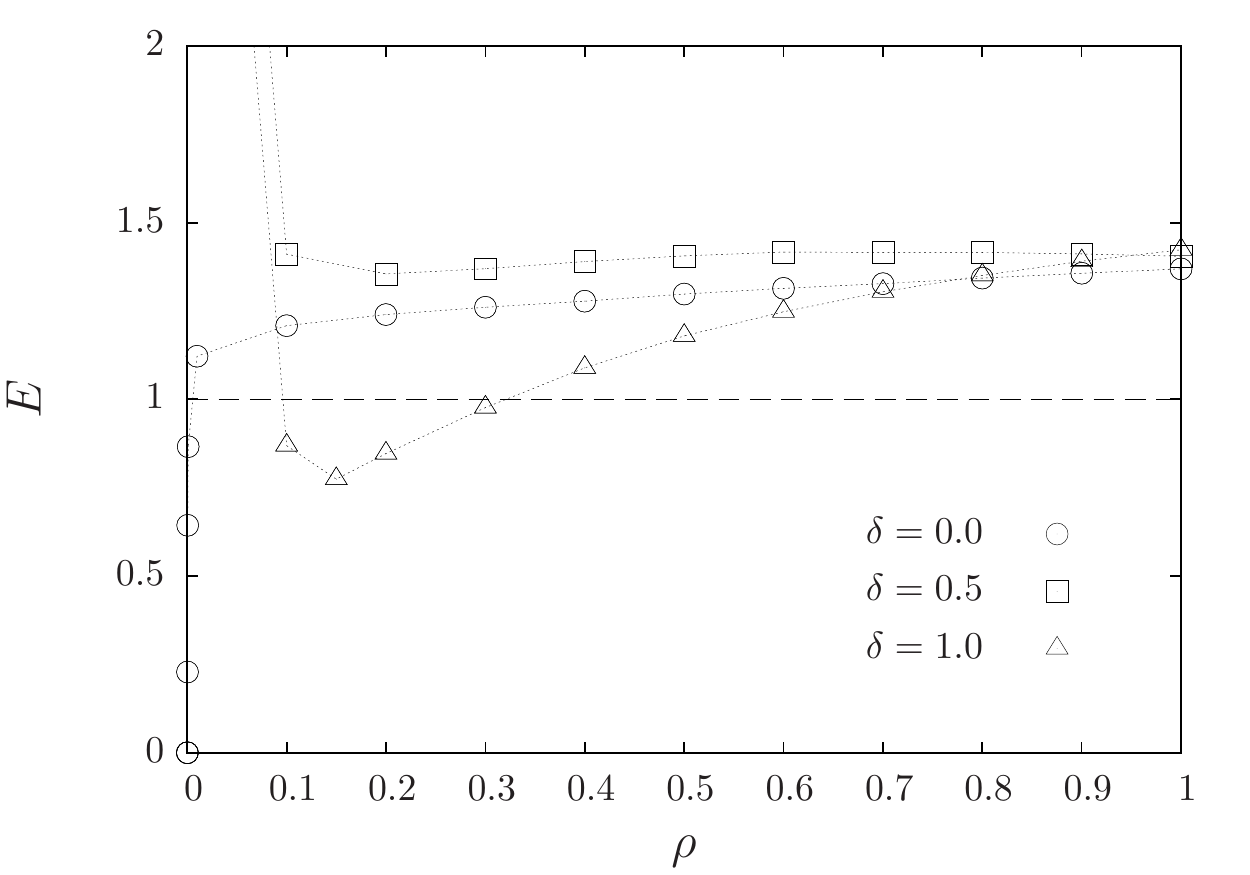}
\caption{Ratio $E\equiv \langle n_\infty \rangle/n_0$ as a function of   the favorable-patch density $\rho$, for $L=100$.
Different values of $\delta$ were also considered as indicated on the figure.  
The remaining parameters are  $A=b=1$,   $D=20$, $\ell_c=0.5$, $\mu=2.0$ and $\sigma_\eta=\sigma_\xi=1$. 
The symbols represent the average over 20 samples and the vertical bars the standard error.
The horizontal line represents $E=1$.   
}
\label{fig:Erho}
\end{figure}

In Fig.~\ref{fig:Erho}, we show the ratio $E$ as a function of $\rho$ for the case $\mu=2$.
By comparing the outcomes for different values of $\delta$, we  see the impact of distinct connectivities. 
In order to interpret this figure,  recall that the initial population density $n_0$ is proportional to the number  
of favorable patches $n_v$, namely $n_0=n_v\, A/b=\rho L^2$.
%
%Moreover,  for  $\rho = 10^{-4}$ and $L=100$,  we have $n_v=1$.  

For $\delta=1$, $E$ presents a minimum value for $\rho \simeq 0.15$. 
Beyond this value, $E$ grows with $\rho$ attaining the value  of the full favorable lattice. 
In the opposite limit of  vanishing $\rho$ (no favorable patches), 
$E$ diverges as far as, according to the model,  (intrinsically) favorable patches are not necessary to promote growth due to the noisy 
growth rate. 
However,  if noise is reduced, then the stochastic dynamics approaches the deterministic one, 
where the population will certainly go extincted.

Now, turning our attention to the $\delta=0$ case, $E$ is monotonically increasing with $\rho$, also attaining a limiting value when $\rho\to 1$. 
Differently from the diffusive case, there exists a critical value  $\rho_c = 4 \times 10^{-4}$ ($n_v=4$) for  population survival.

For small $\rho$, it is curious that the role played by the connectivity, according to the model, 
makes  the diffusive behavior more efficient, while 
selective moves are important at high values of $\rho$.    
In this case,  when the system is approaching a fully favorable landscape, 
the ratio $E$ tends to be the same for different values of $\delta$.   
For intermediate values of $\rho$, we see that the selective strategy overcomes the diffusive one 
(but never overcomes the combined scheme).

\section{Concluding remarks}

We implemented a general model in which the local dynamics, ruled by the canonical model~\cite{canonical}, 
was coupled through different schemes on top of a  complex landscape. 
This setting  allowed to study the role of  the habitat spatial structure and the stochastic fluctuations 
on the long-time state of the metapopulation. 
We restricted the analysis to a  region  of  parameter space relevant to 
display the main features and the interplay between the different processes involved.
For the deterministic case, we have shown that, for small spread rates $D$, 
the distribution of favorable patches must be clustered enough for survival, while below 
a critical value of $\mu$ extinction occurs.
For the stochastic case,  we  have shown that noise in combination with spatial coupling 
has a constructive role, that
drives the population to survival, in contrast to the decoupled case 
where isolated patches would be extincted in finite time. 
We also studied the effects of the spreading strategy,  pointing out that 
a mixed strategy (diffusive dispersion plus selective routes) will result in a larger population size 
(Fig.~\ref{fig:D_mte}). 

Furthermore, we analyzed the ratio $E$ as a function of the coupling range $\ell_c$ and landscape parameter $\mu$ 
for different dispersal strategies. The more clustered, the more viable the environment is. 
For the selective strategy, the coupling range improves $E$, being more effective in disperse environments. 
The effect of the range saturates probably due to the rapid exponential decay of the weight function. 
In contrast, for the diffusive strategy, the coupling range plays and opposite role, as far as it 
drives individuals to unfavorable regions.

Our model could be improved in several directions. 
For instance by considering  correlated environment fluctuations, exhaustible resources,  etc. 
But, despite simple, the model shows the impact of spatial coupling,  spatiotemporal fluctuations and their interplay, 
allowing to foresee the conditions for population survival as well as the optimal dispersal strategy.

%\begin{appendices}
\appendix 
\label{appendix}
% 1 %%%%%%%%%%%%%%%%%%%%%%%%%%%%%%%%%%%%%%%%%%%
\section{Stability of deterministic steady states}
\label{sec:A}

In order to study how steady state stability is affected by spatial coupling, 
let us assume that the population is located at the favorable patches, which is true for small $D$ 
(that is, close to the uncoupled case), and that the coupling is purely diffusive ($\delta=1$). 
For a favorable patch,  the deterministic form of Eq.~(\ref{maineq}) reads
\begin{eqnarray} \nonumber
\dot{u}_i &=& A u_i -b u_i^2 + D\sum_{j\neq i} (u_j-u_i)\gamma(d_{ij}) \\ 
&=& (A-D) u_i -bu_i^2 + D\sum_{j\neq i} u_j\gamma(d_{ij})    \, ,
\label{deteq}
\end{eqnarray}
recalling that $\sum_{j\neq i} \gamma(d_{ij})=1$.
To estimate the last term, that represents the flow of individuals from the neighborhood towards patch $i$, $J_{i}^{in}$,
we consider that   $u_j \approx u_i$. In this case 
\begin{equation}
J_{i}^{in} =   u_i \sum_{j\neq i}  \gamma(d_{ij})\,,  
\label{jin}
\end{equation}
where the sum effectively runs over the $n_v$ favorable patches. 
The average over arrangements of a landscape  $\gamma_\mu \equiv \langle \sum_{j\neq i}  \gamma(d_{ij}) \rangle$ 
can be estimated as  
\begin{equation} 
\label{gammamu}
 \gamma_\mu = n_v \int P_\mu(\ell) e^{-\ell/\ell_c} d\ell \,.
\end{equation}
It depends on $\mu$ and on the density $\rho$, such that it varies from $\rho$ (when $\mu=0$) to 1, in the extreme cases
of either maximal density or very large $\mu$. That is, $\gamma_\mu$ increases with $\mu$, with $\rho$ and  with 
$\ell_c$ too.
 Then, Eq.~(\ref{deteq}) can be approximated by 
\begin{equation}
\dot{u}_i \simeq (A- D[1-\gamma_\mu]) u_i -bu_i^2 \equiv G u_i  -bu_i^2\, .
\label{deteq2}
\end{equation}
If $G >0$,  the population  will grow and assume a finite value,  bounded by the carrying capacity. 
Meanwhile, $D$ diminishes the effective growth rate $G$, that 
becomes negative  for sufficiently large $D$, namely for
\begin{equation}
 D   >A/(1- \gamma_\mu) \, 
\label{growth}
\end{equation}
indicating decrease of the population. 
In fact notice in  Fig.~\ref{fig:det}  that the smaller $D$ the less frequent the extinction events for a given $\mu$.
This effect can be  mitigated by the landscape, 
through parameter $\gamma_\mu$, when   the density of favorable sites or 
clusterization associated with large $\mu$ increases.
Eq.~\ref{growth} also  provides the linear stability condition for the null state. 
If  $G<0$, the population will decrease and go extincted.

\section*{Acknowledgements}
C.A. acknowledges the financial support of Brazilian Research Agencies CNPq  and FAPERJ. 
E.H.C. acknowledges financial support from 
Coordena\c c\~ao de Aperfei\c coamento de Pessoal de N\'ivel Superior (CAPES). 

\bibliographystyle{apsrev4-1}
\bibliography{reference}

\end{document}